%Paper: hep-th/9301048
%From: mokhov%trbilun.BITNET@pucc.Princeton.EDU (Oleg Mokhov)
%Date: Tue, 12 Jan 93 18:00:08 +0200

%Paper: O.I.Mokhov
%  Nonlinear sigma-models
%  and symplectic geometry on loop spaces
%  of (pseudo)Riemannian manifolds

%plain tex
\magnification=\magstep1
\baselineskip=12pt
\font\ab=cmr9
\overfullrule=0pt
\font\twelvebf=cmbx12
\rightline{hep-th/9301048}
\vsize = 7 in
\vglue .7 in
\centerline{\twelvebf  Nonlinear Sigma Models and}
\centerline{\twelvebf  Symplectic Geometry on Loop Spaces of}
\centerline{\twelvebf  (Pseudo)Riemannian Manifolds}
\vskip .7 in
\centerline{\bf Oleg Mokhov\footnote{$^{\dag}$}
{On leave of absence from
{\it VNIIFTRI, Mendeleevo, Moscow Region 141570, Russia}}}
\smallskip
\centerline{\it Department of Mathematics}
\centerline{\it Bilkent University}
\centerline{\it 06533 Bilkent, Ankara, Turkey}
\vskip 1in
\leftskip = 30 pt
\rightskip = 30 pt
\centerline{\bf Abstract}
\smallskip

{\ab In this paper we consider symplectic and Hamiltonian structures of systems
generated by actions of sigma-model type and show that these
systems are naturally connected with specific symplectic geometry
on loop spaces of Riemannian and pseudoRiemannian manifolds.}
\par
\vfill\eject
\hsize = 5 in
\leftskip = 0 pt
\rightskip = 0 pt
\vskip .5in

\noindent{\bf 1  Introduction}

\noindent
We consider two-dimensional systems generated by the sigma-model
actions of the form
$$ S = \int \biggl({1\over 2}g_{ij}(\phi)\phi_x^i\phi_t^j
+ U(\phi) \biggr) dxdt \eqno{(1)}$$
where $g_{ij}(\phi)$ is an arbitrary Riemannian or pseudoRiemannian
metric in a N-dimensional target space M with local coordinates
$ (\phi^1,...,\phi^N)$, the functions $\phi^i(x,t)$ are bosonic fields,
the field $U(\phi)$ is arbitrary here.

It was shown in [1] (see also [2--5]) that any system generated by the
action of the form (1) has always natural Hamiltonian structure given
by explicit symplectic form $ \omega(\xi,\eta) = \int_{\gamma}<\xi,
\nabla_{\dot{\gamma}}\eta>$ on the loop space of the manifold M.
Correspondingly, integrable systems with the action (1), generally
speaking, could be selected (see [2,3]) by description of symplectic structures
(or Hamiltonian structures) compatible with the given symplectic form
$\omega(\xi,\eta)$. In fact, this is the specific problem (see [1--3])
of symplectic geometry on loop spaces. Usually, for integrable systems,
second symplectic structure compatible with the first one (if it exists
there) can be received from another but equivalent action for considered
system.

\vskip.2in

\noindent{\bf 2  Sigma Models and Symplectic Forms on Loop Spaces}

\noindent
Any system generated by the action (1) has the form
$$ g_{ij}(\phi)\phi_{xt}^j + g_{is}(\phi)\Gamma_{jk}^s(\phi)\phi_x^k
\phi_t^j  = {\partial U\over \partial\phi^i} \eqno{(2)}$$
where $ \Gamma_{jk}^s(\phi)$ are the coefficients of the Levi-Civita
connection defined by the metric $g_{ij}(\phi)$ (i.e. the only
symmetric connection compatible with the metric). As it was shown in [1]
the homogeneous matrix differential operator of the first order
$$ M_{ij} = g_{ij}(\phi){d\over dx} + g_{is}(\phi)\Gamma_{jk}^s(\phi)\phi_x^k
\eqno{(3)}$$
is a symplectic operator for any Riemannian or pseudoRiemannian metric
 $ g_{ij}(\phi)$. In other words, the inverse operator  $ K^{ij} =
  (M^{-1})^{ij}$, such that $ K^{ij}M_{jk} = \delta_k^i$, gives always
the nonlocal Poisson bracket
$$ \{\phi^i(x),\phi^j(y)\} = K^{ij}[\phi(x)] \delta(x - y) \eqno{(4)}$$
and hence we have always the following nonlocal Hamiltonian
representation for the system (2):
$$ \phi_t^i = K^{ij}{\partial U\over \partial \phi^j} \equiv
 \{\phi^i(x),\int U(\phi)dx\} \eqno{(5)}$$
The Poisson bracket (4) is nonlocal and very complicated and it is
much convenient to use  local symplectic representation
$$ M_{ij}\phi_t^j = {\partial U\over \partial\phi^j}$$ and study the
corresponding symplectic geometry. The local symplectic operator (3)
gives (see [1]) a natural local symplectic form $\omega$ on loop
space of the manifold M. Really, we have N-dimensional Riemannian
or pseudoRiemannian manifold M with metric $g_{ij}(\phi)$. Consider
the loop space of the manifold M. It means here the space
$L(M)$ of all smooth parametrized mappings $ \gamma : S^1 \to M,
 \gamma(x) = \{\phi^i(x), x \in S^1 \}$. The tangent space of $L(M)$
in its point (a loop) $\gamma$ consists of the all smooth vector
fields $\xi^i, 1 \leq i \leq N$, defined along the loop $\gamma$.
We denote it here by $T_{\gamma}L(M)$ (we note that
$\xi^i(\gamma(x)) \in T_{\gamma(x)}M$, $\forall x \in S^1$, where
$T_{\gamma(x)}M$ is the tangent space of the manifold M in the
point $\gamma(x)$).

Consider natural bilinear invariant form $\omega$ on the loop
space $L(M)$:
$$ \omega(\xi,\eta) = \int_{\gamma} <\xi,\nabla_{\dot{\gamma}}\eta>
\eqno{(6)}$$
where $\xi, \eta \in T_{\gamma}L(M), <\xi,\eta> = g_{ij}\xi^i\eta^j$
 is the natural scalar product on the tangent space $TM$ of
(pseudo)Riemannian manifold $(M,g_{ij}), \dot{\gamma} = \{\phi_x^i\}$
is the velocity vector of $\gamma(x)$, $\nabla_{\dot{\gamma}}$
is the operator of covariant derivation along the loop $\gamma$
that is generated by the Levi-Civita connection $\Gamma_{jk}^i(\phi)$.
The bilinear form $\omega$ satisfies (see [1]) to the following
necessary identities:
\item{(1)} $\omega(\xi,\eta) = - \omega(\eta,\xi)$ (skew-symmetry),
\item{(2)} $ (d\omega)(\xi,\eta,\zeta) = 0$ (i.e. the 2-form
 $\omega$ is closed).

\noindent
The differential $d$ here is given by some infinite-dimensional
generalization of the usual exterior differential for Lie algebra
of vector fields:
$$ (d\omega)(\xi,\eta,\zeta) \equiv \sum_{(\xi,\eta,\zeta)}
\biggl\{ \int \xi^i{\delta\omega(\eta,\zeta)\over \delta\phi^i}dx +
\omega(\xi,[\eta,\zeta])\biggl\} \eqno{(7)}$$
The sign $\sum_{(\xi,\eta,\zeta)}$ means here that the sum
is taken with respect to the all cyclic permutations of elements
$(\xi,\eta,\zeta)$.

Thus $\omega(\xi,\eta)$ is a infinite-dimensional
symplectic form on the loop space
$L(M)$. It is easy to show that $ \omega(\xi,\eta)$ is generated
by the symplectic operator (3), i.e.
$$ \omega(\xi,\eta) = \int_{S^1} \xi^iM_{ij}\eta^jdx \eqno{(8)}$$
The corresponding symplectic representation (see also [5]) of the
system (2) has the form:
$$ \omega(\delta\phi,\phi_t) = \delta H, \eqno{(9)}$$
$$ H = \int U(\phi)dx,$$
where the relation (9) is valid for arbitrary variations $\delta\phi^i$
of the fields $\phi^i$, $H$ is a functional on the loop space $L(M)$.
Thus any system (2) can be given by the relation (9) on the loop space
$L(M)$.

It must be noted that the bilinear form (6) is degenerated.
The null-space of (6) consists of all vector fields parallel
along loop $\gamma$. In particular, the velocity vector field
${\phi_x^i}$ belongs to the null-space of (6) if and only if
${\gamma}$ is a geodesic loop on the target space M. We can
consider the corresponding factor space $L(M)/G$ that means
that two loops on the target space M are equivalent if they
can be coincide after some parallel translation on the
manifold M. The form (6) is nondegenerated on the space
$L(M)/G$.
\vskip.2in

\noindent{\bf 3  Sigma Models with Nontrivial Torsion and
Symplectic Structures on Loop Spaces}

\noindent
Consider now more general systems generated by the actions of
the form
$$ S = \int \biggl({1 \over 2}a_{ij}(\phi)\phi_x^i\phi_t^j + U(\phi)
\biggr)dxdt \eqno{(10)}$$
where $a_{ij}(\phi)$ is an arbitrary tensor, i.e.
$ a_{ij}(\phi) = g_{ij}(\phi) + f_{ij}(\phi)$, where $g_{ij}(\phi)$
is a symmetric tensor (a metric on manifold M) and $ f_{ij}(\phi)$
is a skew-symmetric tensor on M.

Then the corresponding Lagrangian system has also always symplectic
representation defined by the symplectic form
$$ \omega(\xi,\eta) = \int_{\gamma} <\xi,\nabla_{\dot{\gamma}}\eta>
\eqno{(11)}$$
where $ <\xi,\eta> = g_{ij}\xi^i\eta^j, \nabla_{\dot{\gamma}}$
 is a covariant derivation generated by some differential geometric
connection $ \Gamma_{jk}^i(\phi)$ with nontrivial torsion. It was
shown in [1] that if $ \det g_{ij}(\phi) \neq 0$ then the form (11)
is symplectic if and only if the differential geometric connection
$ \Gamma_{jk}^i(\phi)$ satisfies to the following conditions:
\item{(1)} the connection $ \Gamma_{jk}^i(\phi)$ is compatible with
the metric $ g_{ij}(\phi)$, i.e.
$$ \nabla_{k}g_{ij} \equiv {\partial g_{ij} \over \partial \phi^k} -
g_{is}(\phi)\Gamma_{jk}^s(\phi) - g_{js}(\phi)\Gamma_{ik}^s(\phi) = 0;$$
\item{(2)} the torsion tensor $ T_{ijk}(\phi) = g_{is}(\phi)T_{jk}^s(\phi),$
$T_{jk}^i(\phi) \equiv \Gamma_{jk}^i(\phi) - \Gamma_{kj}^i(\phi)$,
with respect to all indices and its gradient (or exterior differential
of the corresponding exterior differential form) vanishes:
$ (dT)_{ijkm} = 0$, in other words, the torsion tensor $T_{ijk}(\phi)$
defines a closed 3-form on the manifold M.

\noindent
Any closed 3-form on the (pseudo)Riemannian manifold $ (M,g_{ij})$
gives the only compatible with the metric $ g_{ij}(\phi)$
differential geometric connection $ \Gamma_{jk}^i(\phi)$ with
nontrivial torsion tensor determined explicitly by this closed
3-form and correspondingly any closed 3-form on M generates
a symplectic form (11) with nontrivial torsion on $L(M)$.
The corresponding symplectic operator with nontrivial torsion
have the form
$$ M_{ij} = g_{ij}(\phi) {d \over dx} + g_{is}(\phi)\Gamma_{jk}^s
(\phi)\phi_x^k   \eqno{(12)}$$
where locally we have
$$ \Gamma_{jk}^i(\phi) = {1 \over 2}g^{is}(\phi) \biggl({\partial g_{sk}
\over \partial \phi^j} + {\partial g_{js} \over \partial \phi^k}
- {\partial g_{jk} \over \partial \phi^s} + T_{sjk}(\phi) \biggr),$$
$$ T_{ijk}(\phi) = {1 \over 2} \biggl( {\partial f_{ij} \over \partial \phi^k}
+ {\partial f_{jk} \over \partial \phi^i} + {\partial f_{ki} \over
 \partial \phi^j} \biggr),  \eqno{(14)}$$
$f_{ij}(\phi) = - f_{ji}(\phi),$   $T_{ijk}(\phi) = {1 \over 2} (df)_{ijk}$.

The Hamiltonian (symplectic) operators compatible with (12) and (3)
will be considered in another publications.

\vskip .6in

\noindent{\bf References}
\vskip.2in
\noindent
\item{[1]} O.I.Mokhov, Symplectic forms on loop space and Riemannian
             geometry, {\it Funct. Anal. Appl.} {\bf 24} (1990)
\item{[2]} O.I.Mokhov, Symplectic forms on loop spaces of Riemannian
             manifolds, {\it Report at the International Conference
             "Differential Equations and Related Problems" in honour
             of 90-th anniversary I.G.Petrovsky (1901-1973), Moscow
             State University, Moscow, USSR, May 1991. Russian Math.
             Surv.} {\bf 46} (1991)
\item{[3]} O.I.Mokhov, Symplectic geometry on loop spaces of smooth
             manifolds and nonlinear systems, {\it Report at the
             3rd International Workshop "Theory of Nonlinear Waves",
             Kaliningrad University, Kaliningrad, USSR, September
             1991}
\item{[4]} I.Ya.Dorfman, O.I.Mokhov, Local symplectic operators
             and structures related to them, {\it Journal Math.
             Physics} {\bf 32} (1991) 3288
\item{[5]} O.I.Mokhov, Two-dimensional $ \sigma $-models in the
             field theory: symplectic approach, {\it Proceedings
             of the 9th Workshop "Modern Group Analysis. Methods
             and Applications", Nizhniy Novgorod, Russia, June
             1992}
\item{[6]} O.I.Mokhov, Two-dimensional nonlinear sigma models
             and symplectic geometry on loop spaces of
             (pseudo)Riemannian manifolds, {\it Report at the
             8th International Workshop on Nonlinear Evolution
             Equations and Dynamical Systems (NEEDS'92), Dubna,
             Russia, July 1992}

\bye